\documentclass[aps,prl,twocolumn,showpacs,superscriptaddress]{revtex4-1}

\usepackage{graphicx}
\usepackage{hyperref}
\usepackage{amsmath}
\usepackage{bm}
\hypersetup{
setpagesize=false,
 bookmarksnumbered=true,%
 bookmarksopen=true,%
 colorlinks=true,%
 linkcolor=blue,
 citecolor=blue,
 urlcolor=black,
}

\begin{document}

\title{Optical Levitation of a Mirror for Reaching the Standard Quantum Limit}
\author{Yuta Michimura}
  \email{michimura@granite.phys.s.u-tokyo.ac.jp}
  \affiliation{Department of Physics, University of Tokyo, Bunkyo, Tokyo 113-0033, Japan}
\author{Yuya Kuwahara}
 \email{kuwahara@granite.phys.s.u-tokyo.ac.jp}
  \affiliation{Department of Physics, University of Tokyo, Bunkyo, Tokyo 113-0033, Japan}
\author{Takafumi Ushiba}
  \affiliation{Department of Physics, University of Tokyo, Bunkyo, Tokyo 113-0033, Japan}
\author{Nobuyuki Matsumoto}
  \affiliation{Frontier Research Institute for Interdisciplinary Sciences, Tohoku University, Aoba, Sendai 980-8578, Japan}
  \affiliation{Research Institute of Electrical Communication, Tohoku University, Aoba, Sendai 980-8577, Japan}
  \affiliation{JST, PRESTO, Kawaguchi, Saitama 332-0012, Japan}
\author{Masaki Ando}
  \affiliation{Department of Physics, University of Tokyo, Bunkyo, Tokyo 113-0033, Japan}
  \affiliation{National Astronomical Observatory of Japan, Mitaka, Tokyo 181-8588, Japan}
  \affiliation{Research center for the early universe, Graduate School of Science, University of Tokyo, Bunkyo, Tokyo 113-0033, Japan}

\date{\today}

\begin{abstract}
We propose a new method to optically levitate a macroscopic mirror with two vertical Fabry-P{\'e}rot cavities linearly aligned. This configuration gives the simplest possible optical levitation in which the number of laser beams used is the minimum of two. We demonstrate that reaching the standard quantum limit (SQL) of a displacement measurement with our system is feasible with current technology. The cavity geometry and the levitated mirror parameters are designed to ensure that the Brownian vibration of the mirror surface is smaller than the SQL. Our scheme provides a promising tool for testing macroscopic quantum mechanics.
\end{abstract}


\maketitle

{\it Introduction.}---
Position measurement of a macroscopic mechanical oscillator in the quantum regime is essential 
for obtaining insights into macroscopic quantum mechanics (MQM).
Moreover such a quantum oscillator should be prepared over various scales to investigate the classical-quantum boundary.
In particular, mass may play a critical role in generating such a boundary, 
mainly motivated by the gravitational decoherence \cite{diosi1989, penrose1996} 
among decoherence models \cite{bassi2013}.
In order to make an oscillator quantum for testing MQM \cite{yanbei2013, arndt2014, marshall2003, hoff2016, muller2008}, 
all classical noises are required to be less than the standard quantum limit (SQL),
the limit of a continuous position measurement \cite{braginsky1996}.

So far, in terms of the mass scale, the SQL at the mechanical resonance and quantum ground-state cooling
have been achieved below the microgram scale \cite{chan2011, teufel2011, peterson2016}.
At the heavier mass scales, reaching the SQL in the free-mass regime, 
which was studied originally for improving the sensitivity of gravitational-wave detectors,
is of interest for generating macroscopic entanglement \cite{muller2008}.
At the gram \cite{corbitt2007a} and kilogram scale \cite{abbott2016a}, 
there have been studies approaching the SQL.
At the microgram and milligram scales, however, 
experiments using small oscillators\cite{bawai2015} and a pendulum \cite{matsumoto2015} are still far from the SQL 
either at the resonance or in the free-mass regime.
For a pendulum, suspension thermal noise is one of the main obstacles.
Thus optical support of an object \cite{ashkin1970, ashkin1971}
instead of mechanical support would be promising
because light does not introduce thermal noise.
For a mirror, the ways of optical levitation have been proposed at the nanogram \cite{singh2010} 
and sub-milligram scale \cite{guccione2013}, 
yet its realization has not been reported at any scales including around the milligram scale.

In this paper, we propose a new configuration of optical levitation of a macroscopic mirror 
with a mass of sub-milligram. 
This enables reaching the SQL in the free-mass regime with respect to the vertical displacement.
We consider a {\it sandwich configuration} with two Fabry-P{\'e}rot cavities that are aligned linearly and vertically.
While using the optical cavities to vertically support the mirror has been previously proposed
\cite{singh2010, guccione2013},
the difference is in the way used to trap the mirror horizontally.
We use optical restoring forces derived from the geometry of our configuration 
instead of optical tweezers \cite{singh2010} or three pairs of double optical spring \cite{guccione2013},
which results in the minimum number of laser beams.
The levitated mirror using our scheme will be applicable to the MQM test proposed in Ref. \cite{muller2008}
at the milligram scale, and also be helpful to the study of reduction of quantum noise in gravitational-wave detectors.

\begin{figure}
\centering
 \includegraphics[width=\hsize]{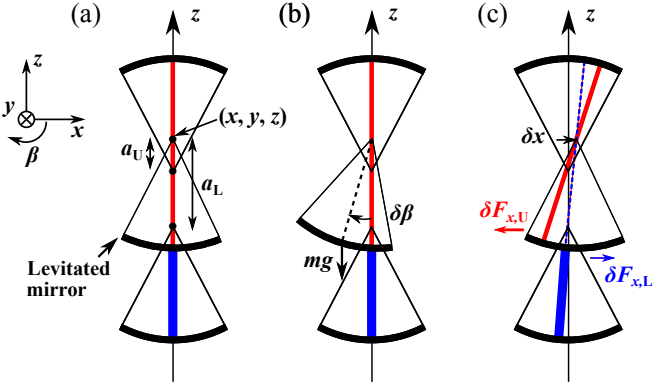}
  \caption{Schematic of the sandwich configuration. 
  The arcs and the centers of the sectors are the mirrors and their COCs, respectively.
  The blue and red lines are the circulating beams in the lower and upper cavities, respectively.
  (a) Balanced state. The vertical motion is constrained by the double optical spring.
  (b) Restoring torque due to the gravity.
  (c) Horizontal restoring force due to the upper cavity. }
\label{sandwich}
 \end{figure}

{\it Sandwich configuration.}---
The sandwich configuration that we propose is composed of two cavities with three mirrors, shown in Fig.~\ref{sandwich}.
At the center is the levitated mirror that is convex downward for the stability of its rotational motion,
with the high reflective (HR) coating at the lower surface. 
At the upper and lower sides are fixed mirrors, each of which forms a Fabry-P{\'e}rot cavity with the common center mirror.
The center mirror is supported by the radiation force from the lower cavity and is stabilized by the upper cavity 
for its horizontal motion. 
The relative positions of the center of curvature (COC) of the three mirrors 
is critical for the optical and mechanical stability of the system 
and must be arranged as shown in Fig.~\ref{sandwich}(a).

A double optical spring \cite{aspelmeyer2014, corbitt2007b} is introduced 
into the cavities for vertical optical trapping without giving rise to suspension thermal noise.
This requires at least two laser beams; one is blue-detuned and the other is red-detuned from the cavity resonance.
Although levitation by using a single cavity is unstable, as explained in the next section, 
the sandwich configuration with two cavities is stable and 
allows for one cavity to be red detuned and the other to be blue detuned.
Thus the sandwich configuration can be realized with the minimum two beams required by a double optical spring.
The frequency of one laser beam compared with the other is shifted by an amount
larger than the cavity linewidth so that each cavity can be regarded as being independent.

{\it Stability.}---
The motion of the levitated mirror is described in terms of 6 degrees of freedom;
the position $(x,\,y,\,z)$ of the COC of the mirror
and the Euler angle $(\alpha,\,\beta,\,\gamma)$.
Here the $z$-axis is set to be the vertical direction. 
A rotational symmetry about the $z$-axis of the system allows us to 
set $\alpha=\gamma=0$ without loss of generality. 
Also, it allows us to confine the motion within the $xz$ plane, equivalent to setting $y=0$.
In this case, the stability of the mirror can be discussed 
by considering the mechanical response to small displacements $(\delta x,\,\delta z,\,\delta \beta)$ 
from its balanced state.
The linear mechanical response is written as 
${^t(}\delta F_x,\,\delta F_z,\,\delta N_\beta) =-K\,{^t(\delta x,\,\delta z,\,\delta \beta)}$, 
where $\delta F_{x,\,z}$ is the restoring force in the $x$ and $z$ direction,
$\delta N_\beta$ is the restoring torque, and $K$ is the $3\times3$ complex matrix 
with its real and imaginary parts representing the spring and damping effect, respectively.
The condition for stability is that both the real and imaginary part of the three eigenvalues of $K$ are positive.

The vertical displacement $\delta z$ and the rotation $\delta \beta$ do not change the positions of the beam spots 
on the center mirror and do not couple to the other degrees of freedom.
The center mirror is subjected to the force
$\delta F_{z}=-(K_\mathrm{L}^\mathrm{opt}+K_\mathrm{U}^\mathrm{opt})\delta z$
due to the double optical spring of the two cavities,
where $K_I^\mathrm{opt}=k_I^\mathrm{opt}+im\omega \gamma_I^\mathrm{opt}$, $m$ is the mass of the mirror,
$\omega/(2\pi)$ is the Fourier frequency,
$k_I^\mathrm{opt}$ is the optical spring constant and  $\gamma_I^\mathrm{opt}$ is the optical damping rate
with $I=\mathrm{L,\,U}$ indicating the lower and upper cavities, respectively.
The stability condition is that 
$k_\mathrm{L}^\mathrm{opt}+k_\mathrm{U}^\mathrm{opt}>0$ and
$\gamma_\mathrm{L}^\mathrm{opt}+\gamma_\mathrm{U}^\mathrm{opt}>0$,
which can be satisfied by adjusting the laser detuning from the cavity resonance.
The mirror is, in turn, subjected to the torque $\delta N_{\beta}=-mgR \cdot \delta \beta$ due to gravity,
as shown in Fig.~\ref{sandwich}(b).
Here $g$ is the gravitational acceleration, and $R$ is the radius of curvature (ROC) of the center mirror, 
the sign of which is defined to be positive when it is convex downward.
To use the gravity as the restoring torque, the mirror must be convex downward, i.e.,  $R>0$.
The rotational motion would be damped by the residual gas instead of the lossless gravity.

On the other hand, the horizontal fluctuation of $\delta x$ changes the positions of the beam spots,
which could lead to an anti-spring effect resulting in the instability of the cavities.
Due to the beam passing through the COCs, 
the direction of the radiation force with the magnitude of $F_I$ tilts by the angle 
$\theta_I$ such that $\tan \theta_I=\delta x/a_I$,
where $a_I$ is the distance between the COCs of the mirrors of the cavity and is shown in Fig.~\ref{sandwich}(a).
This yields the horizontal force 
$\delta F_{x,\,I}=F_I \sin \theta_I \simeq -K_I^\mathrm{hor}\delta x$ 
which includes the damping effect \cite{solimeno1991}, where
\begin{eqnarray}
K_I^\mathrm{hor}=k_I^\mathrm{hor}+i\omega \gamma_I^\mathrm{hor} =
\pm \frac{F_I}{a_I}\left[1-i\omega\frac{\pi l_I}{\mathcal{F}_Ic(1-G_I)}\right],
\end{eqnarray}
$l_I$ is the cavity length, $\mathcal{F}_I$ is the cavity finesse, $c$ is the speed of light, 
$G_I=(1-l_I/R_I)(1-l_I/R)$ and $R_I$ is the ROC of the fixed mirror. 
The sign of $K_I^\mathrm{hor}$ is negative for the lower cavity and positive for the upper cavity, which are determined
by the sign of $\delta F_{x,\,I}$ in Fig.~\ref{sandwich}(c).
The total horizontal force is therefore given by 
$\delta F_{x}=-(K_\mathrm{L}^\mathrm{hor}+K_\mathrm{U}^\mathrm{hor})\delta x$, 
where the real part of the spring constant is $F_\mathrm{U}/a_\mathrm{U}-F_\mathrm{L}/a_\mathrm{L}$. 

\begin{table}
\caption{Parameters for reaching the SQL.
The suffix indicates $\mathrm{s}$ for the substrate, 
$\mathrm{Ta}$ for the $\mathrm{TiO_2}$:$\mathrm{Ta_2O_5}$ coating layer,
$\mathrm{Si}$ for the $\mathrm{SiO_2}$ coating layer,
$\mathrm{L}$ for the lower cavity and $\mathrm{U}$ for the upper cavity.}
\label{parameters}
\centering
\begin{tabular}{lcr} \hline
Levitated mirror &&\\
mass &  $m$ &$0.2\,\mathrm{mg}$  \\ 
radius  & $r$ & $0.35\,\mathrm{mm}$  \\ 
ROC  & $R$ & $30\,\mathrm{mm}$  \\ 
beam radius &  $w_{\mathrm{L,\,U}}$ &$0.14\,\mathrm{mm},\,0.19\,\mathrm{mm}$  \\ 
coating thickness &  $d_\mathrm{Ta}$ & $91\,\mathrm{nm}\times 7\,\mathrm{layers}$ \\
 &  $d_\mathrm{Si}$ & $237\,\mathrm{nm}\times 6\,\mathrm{layers}$ \\
Young's modulus &  $Y_\mathrm{s,\,Ta,\,Si}$ & $73\,\mathrm{GPa},\,140\,\mathrm{GPa},\,73\,\mathrm{GPa}$ \\
Poisson ratio  & $\nu_\mathrm{s,\,Ta,\,Si}$ & $0.17,\,0.28,\,0.17$ \\
loss angle &  $\phi_\mathrm{s,\,Ta,\,Si}$ & $1\times 10^{-6},\,2\times 10^{-4},\,5\times 10^{-5}$ \\ 
refractive index &  $n_\mathrm{s,\,Ta,\,Si}$ & $1.45,\,2.07,\,1.45$ \\ \hline
Laser &&\\
wavelength & $\lambda$ & $1064\,\mathrm{nm}$  \\ 
input power &$P^\mathrm{\,in}_\mathrm{L,\,U}$ & $13\,\mathrm{W},\,4\,\mathrm{W}$  \\ 
frequency noise &$\delta f_\mathrm{a}$ & $0.1\,\mathrm{mHz/\sqrt{Hz}}$  \\  \hline
Cavity &&\\
length &$l_\mathrm{L,\,U}$ & $95\,\mathrm{mm}$,  $50\,\mathrm{mm}$  \\
fixed mirror's ROC & $R_\mathrm{L,\,U}$ & $120\,\mathrm{mm},\,30\,\mathrm{mm}$  \\
COC distance& $a_\mathrm{L,\,U}$ & $5.0\,\mathrm{mm},\,1.3\,\mathrm{mm}$  \\
finesse & $\mathcal{F}_\mathrm{L,\,U}$ & $100$, $100$  \\
intracavity power &$P^\mathrm{\,circ}_\mathrm{L,\,U}$ & $420\,\mathrm{W},\,130\,\mathrm{W}$  \\ 
normalized detuning & $\delta_\mathrm{L,\,U}$ & $-0.005,\,0.018$  \\ \hline
Temperature & $T$ &   $300\,\mathrm{K}$  \\
Air pressure & $P$ &  $10^{-5}\,\mathrm{Pa}$  \\ \hline
\end{tabular}
\end{table}

To summarize the above, the matrix $K$ is diagonalized:
\begin{eqnarray}
K=\left(
\begin{array}{ccc}
K_\mathrm{L}^\mathrm{hor}+K_\mathrm{U}^\mathrm{hor} & 0 & 0 
\\ 0 & K_\mathrm{L}^\mathrm{opt}+K_\mathrm{U}^\mathrm{opt} & 0 \\   
0 & 0 & mgR
\end{array}
\right).
\end{eqnarray}
The condition for stability is that the first and second diagonal components are positive for their real and imaginary parts,
and that for the third component the levitated mirror is convex downward, i.e., $R>0$.

The second order effects are discussed to clarify the trapping range. 
The change of the cavity length $\delta z$ must be much smaller than the length detuning due to the optical spring,
i.e., $|\delta z| \ll \lambda |\delta_I|/\mathcal{F}_I$,
where $\lambda$ is the wavelength of the laser and $\delta_I$ is the laser detuning 
normalized by the (amplitude) cavity decay rate.
With regards to the horizontal range, three effects below could arise. 
First, the horizontal displacement $\delta x$ also changes the cavity length by 
$[a_I^2+(\delta x)^2]^{1/2}-a_I \simeq (a_I/2)(\delta x/a_I)^2$.
Thus $\delta x$ is required to be $|\delta x| \ll (2a_I\lambda |\delta_I| /\mathcal{F}_I)^{1/2}$.
Second, the vertical components of the radiation forces change as
$\delta F_{z,\,I}/F_I=\cos \theta_I \simeq 1-(\delta x/a_I)^2/2$,
and are negligible when $|\delta x| \ll a_I$.
Third, the misalignment between the incident mode and the cavity mode causes the decrease in the intracavity power.
Because the displacement of the beam waist of the cavity mode is approximately $\delta x$,
the mode matching ratio decreases by $(\delta x/w_0)^2/2+(\delta \theta/\theta_0)^2/2$, 
which must be much lower than unity.
Here $w_0$ is the beam waist radius of the cavity mode, 
$\delta \theta\,(=\delta x/a)$ is the tilt of the beam and $\theta_0=\lambda/(\pi w_0)$.
Among the three conditions, the first one would be the most stringent.  

{\it Reaching SQL.}---
We focus on the vertical displacement of an optically levitated mirror to discuss the possibility of reaching the SQL.
The mass $m$ of the mirror determines its SQL in its free-mass regime as
$G_\mathrm{SQL}(f)=2\hbar/(m \omega^2)$, 
where $\hbar$ is the reduced Planck constant and $f=\omega/(2\pi)$ is the Fourier frequency.
Note that in this paper we use single-sideband power spectra of the displacement including $G_\mathrm{SQL}(f)$.
The SQL reaching frequency $f_\mathrm{SQL}$, where the shot noise and the radiation pressure noise intersect,
is determined by the power of the laser used for sensing the displacement \cite{aspelmeyer2014}.
However, $m$ and $f_\mathrm{SQL}$ are not free parameters for the optical levitation 
because the mirror is supported by radiation pressure, namely $mg \approx 2P^\mathrm{\,circ}_\mathrm{L}/c$, 
where $P^\mathrm{\,circ}_\mathrm{L}$ is the intracavity power of the lower cavity.
In the case that $f_\mathrm{SQL}$ is lower than the cavity pole,
$f_\mathrm{SQL}$ is approximately given by
\begin{eqnarray}
f_\mathrm{SQL} \approx \frac{1}{2\pi}\sqrt{\frac{16g}{\lambda}\mathcal{F}_\mathrm{L}},
\label{fSQL}
\end{eqnarray}
which is independent of $m$.

In order to reach the SQL, 
all classical noises must be below the SQL at $f=f_\mathrm{SQL}$.
In our case, the Brownian vibration of the mirror is the most critical among fundamental thermal noises.
Therefore close attention must be paid to the levitated mirror including its material and shape.
In particular, the aspect ratio of the mirror, which we define as diameter over thickness, 
is important and should be close to unity 
because a thinner mirror leads to a lower first resonant frequency and 
thus higher noise level of the Brownian vibration.
In the case that the aspect ratio is close to unity and 
the displacement is sensed at the center of the mirror, 
its noise power spectrum below the mechanical resonant frequency of the mirror is given by 
\cite{levin1998, harry2002, villar2010}
\begin{eqnarray}
&&\frac{4k_\mathrm{B}T}{\omega} 
\left[ \frac{\phi_\mathrm{s}}{\sqrt{\pi} w_\mathrm{L}}\frac{1-\nu_\mathrm{s}^2}{Y_\mathrm{s}}
+ \right. \notag \\
&&\left. \sum_\mathrm{c} \frac{d_\mathrm{c}\phi_\mathrm{c}}{\pi{w_\mathrm{L}}^2}
 \frac{{Y_\mathrm{c}}^2(1+\nu_\mathrm{s})^2(1-2\nu_\mathrm{s})^2
+{Y_\mathrm{s}}^2(1+\nu_\mathrm{c})^2(1-2\nu_\mathrm{c})}
{{Y_\mathrm{s}}^2Y_\mathrm{c}(1-\nu_\mathrm{c}^2)} \right],
\label{coatingTN}
\end{eqnarray}
where the summation runs over $\mathrm{c}=\mathrm{Ta},\,\mathrm{Si}$, 
the first term $G_\mathrm{sub}(f)$ and the second term $G_\mathrm{coa}(f)$ 
correspond to the substrate and coating components, respectively.
Here $k_\mathrm{B}$ is the Boltzmann constant 
and the other parameters are shown in Table~\ref{parameters}.
The coating thickness $d_\mathrm{c}$ is related to the reflectivity of the mirror and thus the cavity finesse $\mathcal{F}_I$.
The beam radius on the mirror $w_I$ is limited by the diffraction loss of the mirror, 
which must be smaller than the cavity's total transmittance $2\pi/\mathcal{F}_I$.
The maximal beam radius is thus roughly proportional to the radius of the mirror, or to $m^{1/3}$ 
with the aspect ratio of the mirror kept constant.
We then have 
$G_\mathrm{SQL}(f_\mathrm{SQL})/G_\mathrm{coa}(f_\mathrm{SQL}) \propto m^{-1/3}\mathcal{F}_\mathrm{L}^{-1/2}$, 
indicating that the lower mass and the lower finesse are advantageous for $G_\mathrm{SQL}/G_\mathrm{coa}>1$.

We show an example set of realistic parameters for reaching the SQL 
that satisfies the stability conditions mentioned above, in Table~\ref{parameters}.
A $0.2\,\mathrm{mg}$ mirror is used, with an aspect ratio of $3$, a radius of $0.35\,\mathrm{mm}$ and 
a ROC of $30\,\mathrm{mm}$.
Its substrate and coating are respectively made of fused silica ($\mathrm{SiO_2}$) 
and alternating titania-doped tantala/silica ($\mathrm{TiO_2}$:$\mathrm{Ta_2O_5/SiO_2}$) layers
with optimal thickness \cite{villar2010}, 
which have been proved to have low loss angle values 
and low optical absorption \cite{harry2007, flaminio2010, pinard2016}.
A laser of $\lambda=1064\,\mathrm{nm}$ is adopted,
with input powers of $13\,\mathrm{W}$ and $4\,\mathrm{W}$ for the lower and upper cavities, 
which are blue and red detuned for the double optical spring, respectively.
Cavity finesses are chosen to be $100$, which is relatively small compared with previous proposals,
to increase the ratio $G_\mathrm{SQL}/G_\mathrm{coa}$ as much as possible.
The distances between the COCs are respectively 
$5.0\,\mathrm{mm}$ and $1.3\,\mathrm{mm}$
for the lower and upper cavities, generating a positive spring effect for the horizontal stability.
Here the effective ROC of the levitated mirror for the upper cavity is modified to $R/n_\mathrm{s}$
due to the HR surface being at the lower side, where $n_\mathrm{s}$ is the refractive index of the substrate.
For these parameters, the trapping ranges of the system are calculated to be
$|\delta x| \ll (2a_I\lambda |\delta_I| /\mathcal{F}_I)^{1/2}=0.6\,\mu\mathrm{m}$ 
and $|\delta z| \ll \lambda |\delta_I| /\mathcal{F}_I=50\,\mathrm{pm}$.

\begin{figure}
\centering
 \includegraphics[width=\hsize]{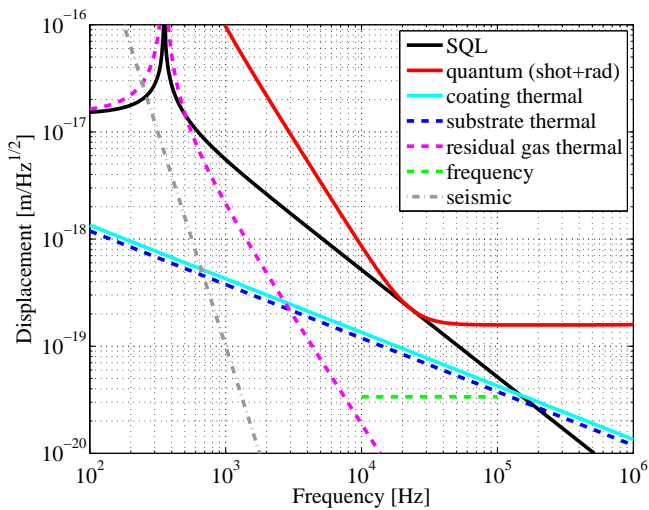}
  \caption{Noise spectra of the $0.2\,\mathrm{mg}$ mirror, using parameters in Table~\ref{parameters}. 
  The mirror reaches the SQL at $23\,\mathrm{kHz}$, 
  corresponding to the vertical displacement of $2.2\times10^{-19}\,\mathrm{m/Hz^{1/2}}$.
  The resonance at $340\,\mathrm{Hz}$ is due to the double optical spring.}
\label{sensitivity}
\end{figure}

For the vertical displacement of the mirror, 
which is measured by using the reflected light of the lower cavity as the change of the lower cavity length,
we obtain the noise power spectra in Fig.~\ref{sensitivity} using the parameters in Table~\ref{parameters}.
The coating Brownian noise is the largest classical source and is lower than the SQL level below $100\,\mathrm{kHz}$.
When other technical noises, which will be discussed below, are reduced to levels smaller than the SQL,
the $0.2\,\mathrm{mg}$ mirror can reach the SQL at $f_\mathrm{SQL}=23\,\mathrm{kHz}$
in its free-mass regime, where the vertical displacement is $2.2\times10^{-19}\,\mathrm{m/Hz^{1/2}}$.
The mirror's mechanical resonances that contribute to the noise in the displacement sensing are 
estimated by finite element analysis to have a lowest frequency of $3.1\,\mathrm{MHz}$.
This is far from $23\,\mathrm{kHz}$, and thus does not affect the thermal noise level there.

{\it Discussion.}---
With regards to other noise sources, the most significant technical noises would be 
classical noises of the laser, consisting of frequency noise and intensity noise.
For a Fabry-P{\'e}rot cavity, the power spectrum of frequency noise is given by 
$G_\mathrm{freq}^{1/2}=l_\mathrm{L}\times \delta f_\mathrm{a}/f_\mathrm{a}$,
where $l_\mathrm{L}$, $f_\mathrm{a}$ and $\delta f_\mathrm{a}$ are the lower cavity length,
the laser frequency and the fluctuation of the laser frequency, respectively. 
The requirement is $\delta f_\mathrm{a}$ to be on the order of $0.1\,\mathrm{mHz/Hz^{1/2}}$ at $23\,\mathrm{kHz}$
to make the frequency noise smaller than the thermal noise level,
which is challenging but within our reach \cite{martynov}.
The classical intensity noise, in turn, must be stabilized at the level of the shot noise.
Although active stabilization is difficult for high input power, 
passive stabilization at the input stage would be possible by using a cavity
with its cavity pole lower than $f_\mathrm{SQL}$ to filter out the intensity noise at $f=f_\mathrm{SQL}$.

A thermal noise will arise due to the viscous damping of the residual gas.
Its damping rate is given by
$\gamma_\mathrm{gas}=SP/(Cm)\times [m_\mathrm{mol}/(k_\mathrm{B}T)]^{1/2}$ \cite{saulson1990},
where $S$, $P$, $C$ and $m_\mathrm{mol}$ are the one-side area of the mirror, 
the air pressure of the residual gas, a constant close to unity depending on the shape of the mirror 
and the mean mass of the gas molecules, respectively.
Assuming that $C=1$, we have $\gamma_\mathrm{gas}=7\times 10^{-8}\,\mathrm{Hz}$
for a vacuum pressure of $P=10^{-5}\,\mathrm{Pa}$, giving a sufficiently small noise level to be negligible.
Seismic vibration noise is known to have the typical spectrum of 
$(10^{-7}/f^2)\,\mathrm{m/Hz^{1/2}}$ above $0.1\,\mathrm{Hz}$ \cite{shoemaker1988}
and would be crucial at low frequency.
However, $f_\mathrm{SQL}$ is so high that it can be fully suppressed at  $f=f_\mathrm{SQL}$ 
using a single suspension of the whole system with its resonant frequency of $1\,\mathrm{Hz}$.

The heat balance of the levitated mirror should be mentioned 
because the extreme increase in temperature worsens the thermal noise level and can possibly deform the levitated mirror.
The mirror receives heat due to the absorption of the beams inside the cavities
as well as environmental radiation (room temperature), and releases heat via its own radiation.
The absorption is dominant for the coating ($\mathrm{TiO_2}$:$\mathrm{Ta_2O_5/SiO_2}$)
, which is $< 0.34\,\mathrm{ppm}$ at $1064\,\mathrm{nm}$ \cite{pinard2016}.
The total absorption is then estimated to be $<0.14\,\mathrm{mW}$,
corresponding to the temperature rise of $<20\,\mathrm{K}$ above the room temperature.
Another issue concerning the laser is that the high intensity of the circulating laser beams could damage the coating.
In our case, the peak intensities are $14\,\mathrm{kW/mm^2}$ (lower light) 
and $2.3\,\mathrm{kW/mm^2}$ (upper light), well within the typical damage threshold of a few $\mathrm{MW/mm^2}$.

{\it Conclusion.}---
We proposed a new configuration, sandwich configuration, 
utilizing optical levitation of a macroscopic mirror for reaching the SQL.
We showed that our configuration gives a stable levitation with the minimum two beams.
Furthermore, we pointed out that the Brownian vibration of the levitated mirror would be the most critical.
We determined a technically feasible parameter set for reaching the SQL with the stability conditions satisfied,
and demonstrated that reaching the SQL with the proposed system is within our reach.

\acknowledgements
We thank Kentaro Komori, Yutaro Enomoto, and Ooi Ching Pin for useful discussions. 
This work was supported by JSPS Grant-in-Aid for Exploratory Research No. 15K13542.



\begin{thebibliography}{99}

\bibitem{diosi1989}
L. Di{\'o}si,
Phys. Rev. A {\bf 40}, 1165 (1989).

\bibitem{penrose1996}
R. Penrose,
Gen. Relativ. Gravit. {\bf 28}, 581 (1996).

\bibitem{bassi2013}
A. Bassi, K. Lochan, S. Satin, T. P. Singh, and H. Ulbricht,
Rev. Mod. Phys. {\bf 85}, 471 (2013).

\bibitem{yanbei2013}
Y. Chen, J. Phys. B {\bf 46}, 104001 (2013).

\bibitem{arndt2014}
M. Arndt, K. Hornberger,
Nat. Phys. {\bf 10}, 271 (2014).

\bibitem{marshall2003}
W. Marshall, C. Simon, R. Penrose, and D. Bouwmeester,
Phys. Rev. Lett. {\bf 91}, 130401 (2003).

\bibitem{hoff2016}
U. B. Hoff, J. Kollath-B{\"o}nig, J. S. Neergaard-Nielsen, and U. L. Andersen,
Phys. Rev. Lett. {\bf 117}, 143601 (2016).

\bibitem{muller2008}
H. M{\"u}ller-Ebhardt, H. Rehbein, R. Schnabel, K. Danzmann, and Y. Chen,
Phys. Rev. Lett. {\bf 100}, 013601 (2008).

\bibitem{braginsky1996}
V. B. Braginsky, F. Y. Khalili, and K. S. Thorne, {\it Quantum Measurement} (Cambridge University Press, Cambridge, UK, 1995).

\bibitem{chan2011}
J. Chan, T. Alegre, A. Safavi-Naeini, J. Hill, A. Krause, S. Groeblacher, M. Aspelmeyer, and O. Painter,
Nature {\bf 478}, 89 (2011).

\bibitem{teufel2011}
J. D. Teufel, T. Donner, Dale Li, J. W. Harlow, M. S. Allman, K. Cicak, A. J. Sirois, J. D. Whittaker, K. W. Lehnert, and R. W. Simmonds,
Nature {\bf 475}, 359 (2011).

\bibitem{peterson2016}
R. W. Peterson, T. P. Purdy, N. S. Kampel, R. W. Andrews, P.-L. Yu, K. W. Lehnert, and C. A. Regal,
Phys. Rev. Lett. {\bf 116}, 063601 (2016). 

\bibitem{corbitt2007a}
T. Corbitt, C. Wipf, T. Bodiya, D. Ottaway, D. Sigg, N. Smith, S. Whitcomb, and N. Mavalvala,
Phys. Rev. Lett. {\bf 99}, 160801 (2007).

\bibitem{abbott2016a}
LIGO Scientific Collaboration and Virgo Collaboration,
Phys. Rev. Lett. {\bf 116}, 131103 (2016).

\bibitem{bawai2015}
M. Bawaj {\it et al.},
Nat. Commun. {\bf 6}, 7503 (2015).

\bibitem{matsumoto2015}
N. Matsumoto, K. Komori, Y. Michimura, G. Hayase, Y. Aso, and K. Tsubono,
Phys. Rev. A {\bf 92}, 033825 (2015).

\bibitem{ashkin1970}
A. Ashkin, Phys. Rev. Lett. {\bf 24}, 156 (1970).

\bibitem{ashkin1971}
A. Ashkin, J. M. Dziedzic, Appl. Phys. Lett. {\bf 119}, 283 (1971).

\bibitem{singh2010}
S. Singh, G. A. Phelps, D. S. Goldbaum, E. M. Wright, and P. Meystre,
Phys. Rev. Lett. {\bf 105}, 213602 (2010).

\bibitem{guccione2013}
G. Guccione, M. Hosseini, S. Adlong, M. T. Johnsson, J. Hope, B. C. Buchler, and P. K. Lam,
Phys. Rev. Lett. {\bf 111}, 183001 (2013).

\bibitem{aspelmeyer2014}
M. Aspelmeyer, T. J. Kippenberg, and F. Marquardt,
Rev. Mod. Phys. {\bf 86}, 1391 (2014).

\bibitem{corbitt2007b}
T. Corbitt, Y. Chen, E. Innerhofer, H. M{\"u}ller-Ebhardt, D. Ottaway, H. Rehbein, D. Sigg, S. Whitcomb, C. Wipf, and N. Mavalvala,
Phys. Rev. Lett. {\bf 98}, 150802 (2007).

\bibitem{solimeno1991}
S. Solimeno, F. Barone, C. de Lisio, L. Di Fiore, L. Milano, and G. Russo,
Phys. Rev. A {\bf 43}, 6227 (1991).

\bibitem{levin1998}
Y. Levin,
Phys. Rev. D {\bf 57}, 659 (1998).

\bibitem{harry2002}
G. M. Harry, A. M. Gretarsson, P. R. Saulson, S. E. Kittelberger, S. D. Penn, W. J. Startin, S. Rowan, M. M. Fejer, D. R. M. Crooks, G. Cagnoli, J. Hough, and N. Nakagawa,
Class. Quantum Grav. {\bf 19}, 897 (2002).

\bibitem{villar2010}
A. E. Villar, E. D. Black, R. DeSalvo, K. G. Libbrecht, C. Michel, N. Morgado, L. Pinard, I. M. Pinto, V. Pierro, V. Galdi, M. Principe, and I. Taurasi,
Phys. Rev. D  {\bf 81}, 122001 (2010).

\bibitem{harry2007}
G. M. Harry, M. R. Abernathy, A. E. Becerra-Toledo, H. Armandula, E. Black, K. Dooley, M. Eichenfield, C. Nwabugwu, A. Villar, D. R. M. Crooks, G. Cagnoli, J. Hough, C. R. How, I. MacLaren, P. Murray, S. Reid, S. Rowan, P. H. Sneddon, M. M Fejer, R. Route, S. D. Penn, P. Ganau, J. M. Mackowski, C. Michel, L. Pinard, and A. Remillieux,
Class. Quantum Grav. {\bf 24}, 405 (2007).

\bibitem{flaminio2010}
R. Flaminio, J. Franc, C. Michel, N. Morgado, L. Pinard, and B. Sassolas,
Class. Quantum Grav. {\bf 27}, 084030 (2010).
 
\bibitem{pinard2016}
L. Pinard, C. Michel, B. Sassolas, L. Balzarini, J. Degallaix, V. Dolique, R. Flaminio, D. Forest, M. Granata, B. Lagrange, N. Straniero, J. Teillon, and G. Cagnoli,
Appl. Opt. {\bf 56}, C11 (2017).
 
\bibitem{martynov}
D. V. Martynov and others,
Phys. Rev. D {\bf 93}, 112004 (2016).
  
\bibitem{saulson1990}
P. R. Saulson,
Phys. Rev. D  {\bf 42}, 2437 (1990).

\bibitem{shoemaker1988}
D. Shoemaker, R. Schilling, L. Schnupp, W. Winkler, K. Maischberger, and A. R{\"u}diger,
Phys. Rev. D {\bf 38}, 423 (1988).

\end{thebibliography}
\end{document}